# Size-induced structural phase transition at ~6.0 nm from mixed fcc-hcp to purely fcc structure in monodispersed nickel nanoparticles

Tarachand[1], Vikash Sharma[1], Jaiveer Singh[2], Chandrani Nayak[3], Dibyendu Bhattacharyya[3], Netram Kaurav[4], Shambhu N. Jha[3] and Gunadhor S. Okram[1]

[1]UGC-DAE Consortium for Scientific Research, University Campus, Khandwa Road, Indore 452001, MP, India

[2] Department of Physics, ISLE, IPS Academy, Rajendra Nagar, Indore 452012, MP, India

[3]Atomic & Molecular Physics Division, Bhabha Atomic Research Centre, Mumbai 400085, Maharastra, India

[4] Department of Physics, Government Holkar Science College, A. B. Road, Indore 452001, MP, India.

E-mail: okramgs@gmail.com and okram@csr.res.in

**Abstract**

We have investigated the core issue of atomic lattices in monodispersed Ni nanoparticles (NPs) of sizes 3.8 nm to 10.1 nm using detailed analysis of X-ray diffraction, synchrotron radiation X-ray absorption spectroscopy (XAS) and magnetization data. This has revealed the very remarkable coexistence of atomic face-centered cubic (fcc) and hexagonal closed-packed (hcp) lattices in samples with particle size ≤ 6.0 nm with the prevalence of only fcc phase beyond this. They are also associated with reduced coordination number, modified electronic structure, and surface atom coordination with ligands. Magnetization data furthermore reveal coexistence of ferromagnetism and superparamagnetism at 300 K. Considered to be due to dominant roles of ligands, they are likely to open up far-reaching implications to their future applications.

Keywords: Crystal structure, structural phase transition, XANES, EXAFS, magnetization.

**1. Introduction**

Advanced future novel detector, optical, magnetic recording and electronic metamaterials [1-6] and fundamental understanding [7,8] of nanoparticles (NPs) will depend substantially on the assembly of NPs into three-dimensional (3D) NP lattice that forms their own nanolattice in addition to the already partially disordered atomic lattice(s) inside each of the NP. This is so since nanolattices are superior [9-11] to the isolated NPs and their bulk counterparts for the basically same composition. In this endeavour, there are several precautions materials scientists need to take care. First, the formation of the nanolattice should be made with the surfactant/s meant for synthesizing the NPs. The reason is that they necessarily influence the properties of the final product. In this direction, earlier believe for over a decade of the requirement of using external forces as their cohesive forces [12-15] for the nanolattice formation might change with the discovery of self-generated cohesive forces in enabling a natural nanolattice formation similar to those of atomic lattices [16,17]. The nanolattice however additionally presents further dimension of complexity to the already complicated NPs.

Secondly, we cannot appreciate them precisely unless we understand fully their significant influence on the synthetic conditions, in addition to their surface, size, shape and hence their assembly. The reasons are their profound sensitivities to the nascent surface dangling bonds that get reacted easily. These are quite often controlled nominally with the surfactant used for their stabilization [12,16-22]. These situations impose directly on their combined atomic and NP crystal structures, and their consequent properties and applications. The commonly known obstacles during, and after, the synthesis of metal NPs is the spontaneous surface oxidation in a wide range of spectrum [12,16-22].

Lastly, choice of particular surfactant/s is/ are essential to enable natural nanolattice formation [16,17]. This was illustrated using trioctylphosphine (TOP) wherein natural bulk 3D hcp nanolattices can be formed in the systematically tuned nickel NPs of different average particle sizes. These hcp nanolattices have c/a ratios analogous to the usual atomic hcp lattice. While their crystal structures in nanoscale are analogous to the atomic lattices, possible peculiar changes in physical properties due to the modified environment of atomic lattice, particle size and surfactant have not yet been explored precisely. These complexities have been investigated here through their in-depth X-ray diffraction (XRD), X-ray absorption spectroscopy (XAS) and magnetization data analysis. There is systematic change in atomic lattice structures and their associated change in empty density of states, local symmetry, coordination number and local structures with the change of particle size. As the particle size increases, there is an orderly evolution of coexistence of atomic hcp and fcc structures to completely fcc structure as revealed from the Rietveld fittings of the XRD. This has been corroborated with XAS and magnetization data. The latter in turn also revealed the coexistence of ferromagnetism and superparamagnetism.

## 2. Experimental details

For these studies, we have used monodispersed Ni nanoparticles of sizes 3.8 nm (Ni31), 4 nm (Ni37), 6.0 nm (Ni38), 7.1 nm (Ni36), 9.0 nm (Ni35) and 10.1 nm (Ni34) prepared by thermal decomposition method (Scheme I (a)) [16]. These samples have already been preliminarily characterized [16,17] by X-ray diffraction (XRD), grazing incident XRD (GIXRD), Fourier-transformed infra-red (FTIR), high resolution/ transmission electron microscopy (HRTEM/ TEM), zeta potential and X-ray photoelectron spectroscopy (XPS). In the present study, more detailed analysis of XRD using Fullprof Rietveld fitting, X-ray absorption near-edge structure (XANES) and extended X-ray absorption fine structure (EXAFS) using standard analytical procedures to investigate the changes in oxidation state and local structure with the variation of size of nanoparticles was carried out. The energy calibration was done by defining the pre-edge peak of the nickel foil XANES as 8333 eV. The magnetic properties were investigated using (Quantum design, USA) superconducting quantum interference device - vibrating sample magnetometer.

## 3. Results and discussion

### 3.1. *X-ray diffraction (XRD) study*
Dominant role of TOP on particle size, monodispersity and natural nanolattice formability was discussed earlier [16,17]. Briefly, as TOP concentration increases from 1 ml to 10 ml (Scheme I (a)), (i)) the particle size determined from transmission electron microscopy (TEM) decreases from 10.1 nm down to 3.8 nm (Scheme I (b) and Figure 1 (a)), and (ii) the peak intensities near $2\theta = 44.3°$, $51.6°$ and $76.2°$ due to (111), (200) and (220) lattice planes of fcc structure gradually diminish such that the first peak becomes broader while the two smaller intensity peaks at higher angles disappear. Finally, only one broad peak at $44.3°$ remains for 3.8 nm sample (Figure 1 (b)). Calculation of particle size using Scherrer formula for spherical (crystalline domain of) nanoparticles [23] showed that corresponding size of the NPs varies between ~1.1 nm and 10.1 nm. It should be pointed out that there is drastic difference in size for ≤ 6.0 nm (TEM) between Scherrer and TEM sizes, say 3.8 nm TEM size corresponds to 1.1 nm Scherrer size. For bigger particles than 6.0 nm, both TEM and Scherrer sizes were same within the experimental error of ± 0.1 nm. We therefore use here the size determined from TEM unless otherwise mentioned. We also point out that in contrast to the reported complete oxidation of Ni nanoparticles, prepared using same surfactants, within 24 h [7], we find negligible oxidation even after several weeks of preparation [16].

Intriguingly, when the particle size is reduced to 3.8 nm, a single peak near 44.5° could be due to either hexagonal closed-packed Ni (011) and fcc Ni (111) peaks together or separately as they are close to 44.5°. This conjecture is in line with the probability to form hcp on/ near the surface of the particle and/ or the whole of the NPs when all the unit cells lie on the surface *only*, as hcp phase being less stable thermodynamically than fcc phase [24]. Since they may not be necessarily intermixed, Ni NPs might have a possible formation of its comparatively less stable hcp structure at/ near the surface or the outer shell, and more stable fcc structure at the core. That is, a typical NP might have a central core (fcc), inner shell (hcp) and outer shell due to cappants, which is shown schematically in Scheme 1 (b).

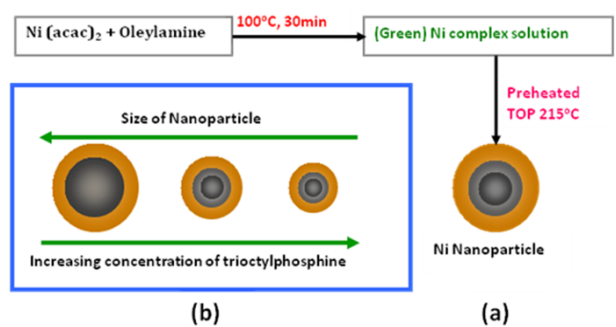

**Scheme 1**(a) Typical synthesis process of Nickel nanoparticles by thermal decomposition method and (b) effect of concentration of TOP on particle size.

Consequently, in spite of the very good surface morphology and high monodispersity of the particle sizes and hence forming their own natural nanolattices [16,17] i.e. lattice of nanoparticles (e.g. Figure 1 (a), for more details, see ref. 16 &17), these NPs are not so straightforward in terms of their atomic lattices. In order to demonstrate this, we have investigated their bulk XRD using Rietveld Fullprof software (Figure S1, Tables S1 in Electronic supporting information). As a typical example, XRD pattern of 3.8 nm sample is shown in figure 1(b). The coexistence of fcc structure with hcp structure has been realized when we Rietveld-fitted XRD data. It was found that fcc structure only could not provide a good fit, rather a combination of them did. This is shown for 3.8 nm sample as a representative XRD (Figure 1 (b)). From the fitting, it was found that fraction of fcc structure was about 3.5% while that of hcp structure was about 96.5%. For 4.0 nm sample, these were 12% and 88%, respectively. As per the energetic [25], say for 3.8 nm, the former has been assigned to the core, the latter to the intermediate shell and the surfactant as the outer shell (Scheme I (b)). If we assume that the core/ crystallite region is identified by Scherrer method (~1.14 nm) and those of shells by TEM image (3.8 nm) minus the former, then they become 2.6% and 97.4%, respectively, matching reasonably well those found from XRD analysis. This may suggest that the shell should consist of *inner* shell due to hcp phase and *outer* shell due to surfactant [16,17]. The mismatch is attributed to the overestimation in TEM size, compared to that of XRD. Further, fcc structure of, say, 3.8 nm sample showed a lattice parameter a = 3.533 Å and nearest-neighbour (NN) bond length r = 2.498 Å while its corresponding bulk values are 3.540 Å and 2.5095 Å, respectively. This shows a reduction of 0.2% from its bulk values. Similarly, its hcp values are a = b = 2.640 Å, c = 4.220 Å with r = 2.640 Å while those for bulk are 2.6515 Å, 4.343 Å and 2.6515 Å, with a corresponding reduction of 0.4%, 2.8% and 0.4%, respectively.

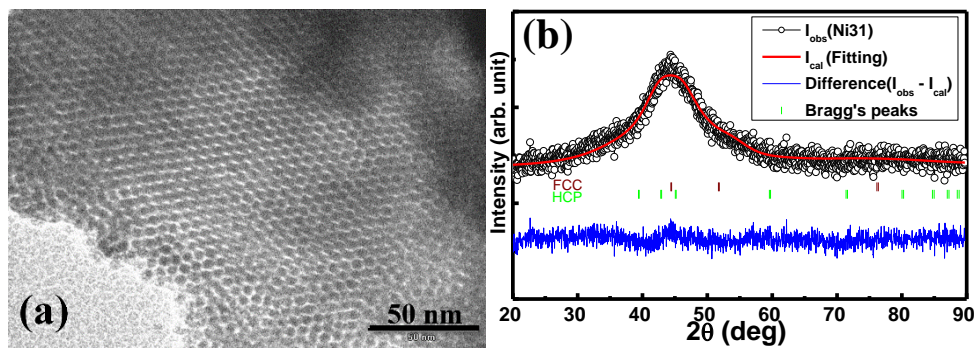

**Figure 1** Representative (a) transmission electron microscopy image and (b) XRD pattern of 3.8 nm nickel nanoparticle sample along with Rietveld fitted (red) curve, difference (blue) curve and Bragg peak positions (vertical lines) of fcc and hcp structure of nickel.

Note that while change in lattice parameter and bond length in fcc structure is quite small, those of hcp are somewhat large. However, as the size increases, both these parameters approach and then surpass that of the bulk (Figure 2 and Table S1). On the other hand, this situation of coexistence of fcc and hcp phases together (i) below 6.0 nm from which onward a pure fcc structure only prevails, (ii) demonstrates highly complex nature of these monodispersed NPs below 6.0 nm and (iii) may lead to rather interesting (physical) properties [25].

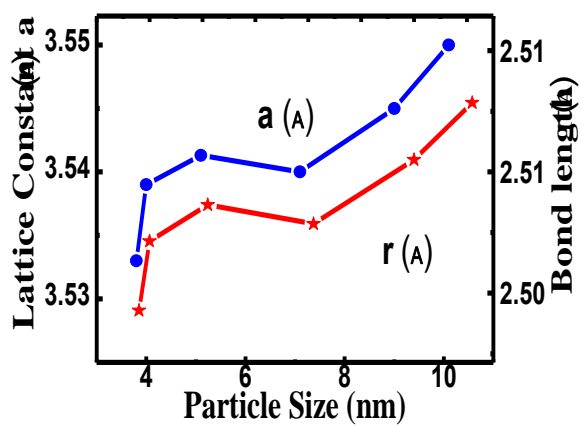

Figure 2 Lattice parameter a and bond length r. Plots of a and r as a function of particle size.

The significant change in lattice parameter and bond lengths implies that the coordination number (CN or N) of Ni atoms would have considerably been disturbed [26]. This is evident from the X-ray absorption near-edge structure (XANES) and extended X-ray absorption fine structure (EXAFS) spectroscopy discussed later. While oxide of Ni is negligible, these NPs on their surface are associated with ligands with/without fully bonding as a compound of nickel [16,17]. Consequently, they are not probably directly associated with the significant oxidation of nickel but some type of covalent bonding [17]. Thus, dominantly octahedral, plus partially tetrahedrally coordinated Ni atoms in their bulk, have suggestively been distorted, reduced site occupancy, modified (virtually) valence, and changed metallic character [27] in line with the XAS data below.

3.2. *XANES study*

In X-ray absorption near-edge structure (XANES) region, ejected photoelectron probes the empty density of electronic states of the materials. As a result, XANES has long been known to be rich in chemical and structural information [27-29]. In order to understand these on Ni NPs, figure 3 shows the XANES spectra of several Ni NP samples of varying particle sizes along with Ni foil and NiO as reference samples, which, in principle, contain 100 % +0 and +2 state of Ni, respectively. Ni foil and Ni NP sample spectra exhibit prominent pre-edge peak P1 due to the 1s→3d transition and main absorption peak P2. NiO does not exhibit (prominent) pre-edge peak but with very intense peak P2, the highest of all.

The positions in the absorption edge peaks P1 and P2 of the NPs are nearly same as that of Ni foil. However, evidence of systematic increase in P1 and decrease in P2 peak intensities as the particle size increases might indicate systematic change in number or type of coordination and symmetry of environment of nickel atoms in NPs. Pre-edge peak P1 due to quadrupole 1s→3d transition is in fact forbidden by dipole selection rule. In bulk nickel, dominantly octahedral plus partially tetrahedrally coordinated Ni atoms allows the 1s→3d transition maximally. This quadrupole transition gradually diminishes as size of NPs reduces.

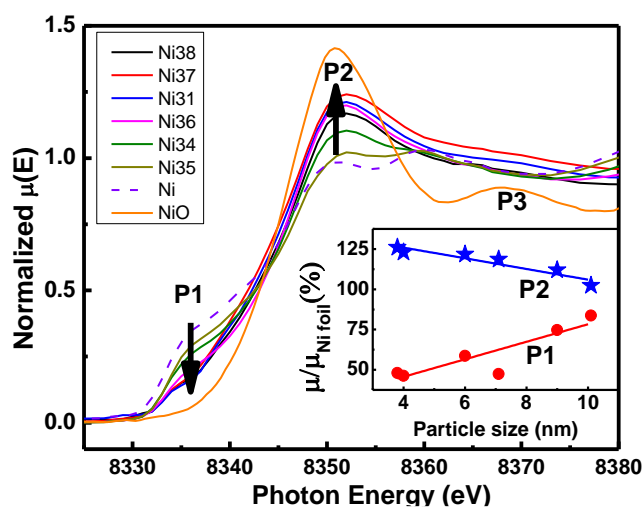

**Figure 3** X-ray absorption near edge spectra. XANES spectra of 3.8 nm (Ni31), 4 nm (Ni37), 6.0 nm (Ni38), 7.1 nm (Ni36), 9.0 nm (Ni35) and 10.1 nm (Ni34) Ni nanoparticle samples along with Ni foil and NiO as reference samples. Arrows indicate the systematic peak height reversal at P1 and P2 as the particle size increases. Inset: plots of how these absorption peak intensities change with particle size. Lines are the best fits.

This therefore gives evolving characteristics of local bonding coordination [30]. This is in agreement with EXAFS data discussed below. This shows modified symmetry of nickel site [31]. Interestingly, while peak P1 intensity increases with increase in particle size that of P2 concurrently decreases. This is illustrated by P1 and P2 arrows, and in figure 3, inset, as plots of ratios of corresponding peak intensities and that of Ni foil. P1 shows increasing trend towards bulk value of 100% while P2 exhibits decreasing trend from about 125% at 3.8 nm samples towards its bulk value of 100% near 10.1 nm NP sample (Figure 3, inset). Reduction in P1 peak implies decrease in 3d empty density of states or reduction in the hybridization of p and d states. This means that a substantial part of Ni-Ni bonds has been reduced as a result of coordination with ligands and or missing Ni lattice points. This is in accordance with EXAFS data, wherein drastic reduction in CN is revealed.

XANES spectra of Ni NPs are thus characterized by (a) white line peaks much weaker than that of NiO but (b) sharper than that in Ni foil spectrum. Further, in correlation with the metallic electrical conductivity of Ni metal compared to semiconducting NiO, one may conclude the scenario of increasing P2/ decreasing P1 as indication of reduction of conduction electrons in the conduction band as particle size decreases, finally tending towards that of NiO with no/ reduced conduction electrons. XANES features of 10.1 nm and 9.0 nm Ni NPs tend towards that of bulk Ni metal spectra rather closely reproducing all three peaks, P1, P2 and peak above P2. As particle size decreases, peak features evolve systematically deviating significantly from that of Ni foil with the disappearance of the peak above P2 and tending somewhat towards that of NiO spectrum, yet not so close to it. Hence, the empty density of states of these nanoparticles is quite different from that of NiO bulk.

Since for K edge absorption i.e. a 1s state, the photoelectron has to go to a p state, even if states are available with the right energy, there might be no 1s absorption if there are no available p states. NiO has two unfilled 3d electrons near the Fermi level, and a filled 3p band while empty 2p electron states of oxygen are too far away to appreciably over-lap with the metal 1s band. Therefore, intense peak at P2 for NiO arises due to hybridization of O 2p and metal 3d levels but a feeble one arises in Ni foil. Positions of pre-edge peaks P1 of the NPs, being matched nearly with that of Ni foil, is correlated to nearly zero oxidation i.e. metallic state of nickel. However, while intensity, and shape, of P1 gradually decreases, those of P2 increases with particle size. They clearly show significantly modified metallic ligand and coordination environment due to the ligands [16]. This enables the formation of empty p-states somewhat similar to that of NiO, yet maintaining their metallic character (as XRD patterns suggest), consistent with Fourier-transformed infrared (FTIR) [17] and X-ray photoelectron spectroscopy (XPS) [16] data.

These spectral features are attributed to the probable enhanced local displacements or missing lattice points of the atoms (see more details in EXAFS section) that are significantly exposed to surface ligands. One plausible reason for enhanced P2 with particle size reduction could be increasing number of surface atoms being coordinated (probably through complexation [16]) with ligands leading to enhanced (3p states) covalent-like bonding of metallic Ni. They mimic somewhat like that of NiO yet limited to the surface Ni atoms without disturbing much to the metallic character and or loss of Ni atomic lattice points in fcc lattice of bulk nickel due to small size and increased surface exposure [16]. This in turn correlates with the formation of both fcc and hcp structures at the very small particle size regime as discussed above. This also is similar to the de-charging effect of the NPs with size reduction through the surfactant in line with those seen in platinum-germanium alloys [32] and their enhanced resistivity.

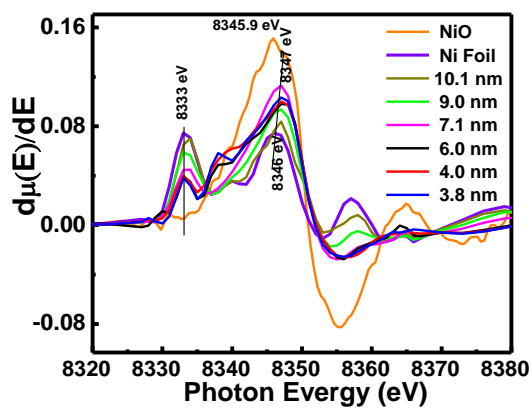

**Figure 4** F Derivatives of the absorption spectra in figure 3 of the samples as indicated.

They turn towards the observed semiconducting nature departing from the original metallic behaviour [33], and similar to those in Co NPs [34]. Thus, XANES data, consistent with their XRD, XPS and FTIR data [16], proved that these Ni NPs are in metallic-like state in bulk but somewhat covalent-like due to the ligands attached to the dominant surface Ni atoms. This does not lead exactly identical to those of NiO bonding and consequently causes distortion or displacement of the (surface) atoms in overall forming a mixture of fcc and hcp structures for NPs at or below 6.0 nm as XRD data analysis indicated.

In order to assess the absorption edge transitions more comprehensively we have taken the derivatives of the XANES spectra in figure 3 and it is shown in figure 4. Peaks near 8336 eV represent the pre-edge transition energy while those near 8346 eV are due to white line absorption edge. These latter peaks tend to shift to higher energy as the particle size decreases indicating that 1s electronic states turn more destabilized requiring more energy to excite or due to quantum size effect [16,26]. The absorption edge of NiO on the other hand leads to a peak near 8345.9 eV. It is thus clear that the absorption spectra of Ni bulk and its nanoparticles are significantly different from that of NiO.

3.3 *EXAFS study*

Figure 5 shows the stacked experimental Ni K edge XANES and EXAFS µ(E) spectra of the Ni NP samples with each spectrum shifted by 0.1 vertically for clarity. Change in the spectra for Ni NPs compared to that Ni foil is quite significant throughout XANES and EXAFS regimes, wherein a significant reduction in EXAFS oscillation amplitudes in Ni NP samples is seen. We list here some of the distinctly evolving features due to size reduction in these samples. Ni foil exhibits two twin peaks at 8351 eV and 8360 eV, six relatively prominent peaks at 8385 eV, 8436 eV, 8508 eV, 8597 eV and 8692 eV and two relatively weaker peaks at 8484 eV and 8556 eV. While these trends are nearly maintained, though the twin peaks at 8351 eV and 8360 eV becoming only one peak near 8353 eV, peak intensities gradually get reduced as the particle size reduces from 10.1 nm to 7.1 nm through 9.0 nm sample. Even though the feeble peaks exist for 6.0 nm sample, they gradually turn to only three main peaks near 8353 eV, 8415 eV and 8496 eV with a hump-like feature near 8367 eV. These three peak features only finally survive in 3.8 nm and 4.0 nm samples. These tendencies are associated with the gradual change in the position of the peaks as the particle size reduces, strongly indicating the gradual change in the environment of Ni atoms, say, coordination number, bond distance and mean square deviation. To ascertain these changes, we have investigated their details using standard EXAFS analysis.

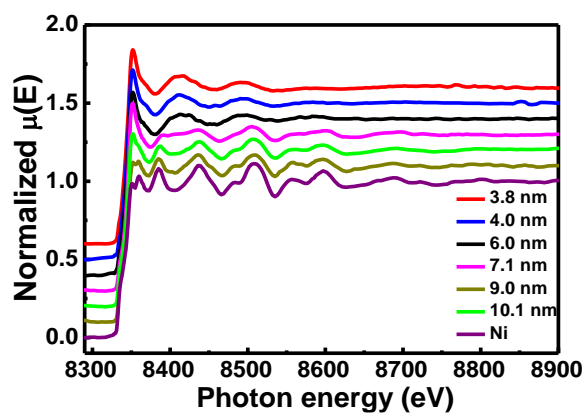

Figure 5 EXAFS spectra of Ni foil, 3.8 nm, 4.0 nm, 6.0 nm, 7.1 nm, 9.0 nm and 10.1 nm nanoparticle samples.

Figure 6 and 7 show the experimental χ(k) spectra weighted with $k^2$ and χ(r) of Ni NP samples. These EXAFS spectra exhibit systematic evolution of the amplitudes of spectral intensity in support of the XANES data shown in Figs. 3 and 5. χ(k) spectra (Figure 6) for 10.1 nm sample show six prominent peaks with a smaller peak. These features are seen in 9.0 nm and 7.1 nm samples though with a slight reduction in their prominence.

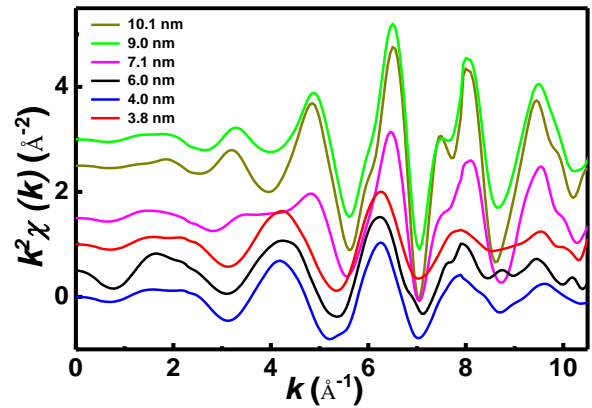

**Figure 6** EXAFS χ (k) spectra weighted with $k^2$ of 3.8 nm, 4.0 nm, 6.0 nm, 7.1 nm, 9.0 nm and 10.1 nm nanoparticle samples.

They are drastically reduced in 3.8 nm, 4.0 nm and 6.0 nm samples with shifted peak positions suggesting their local structure changes compared to those of larger particle size samples. This is clear from figure 7, wherein the first peak positions near 2.0 Å for the former three larger particle size samples while it is seen near 1.626 Å for the latter three smaller size particle samples. For 10.1 nm, 9.0 nm and 7.1 nm samples, first peak comes near 2 Å (without phase correction). However, for 3.8 nm, 4.0 nm and 6.0 nm samples, peak shifts towards lower values near 1.5 Å (without phase correction). That is, for particle size < 6.0 nm, peak position lies between Ni-O peak of NiO structure and Ni-Ni peak of Ni fcc structure. This suggests a change in structure as particle size decreases below 6.0 nm as has also been observed from XRD data analysis and in XANES spectra of these samples described above. It was realized during the EXAFS fitting process that it was not possible to resolve the NP samples with a model that included only a pure, well-ordered fcc Ni crystal structure. While the experimental EXAFS spectra of 10.1 nm and 9.0 nm samples have been fitted with theoretical spectra assuming Ni-Ni path at 2.50(12×) Å of Ni fcc structure, 7.1 nm sample peak shows broadening and could not be fitted assuming single Ni-Ni shell. Therefore, this sample has been fitted from 1.5 to 3 Å assuming a Ni-O shell at 2.08(6×) Å and a Ni-Ni shell of Ni fcc phase at 2.50(12×) Å. From XRD results, it is already known that the samples with size < 6.0 nm have both Ni hcp and fcc phases. Ni hcp structure consists of two Ni-Ni shells at 2.31(6×) Å and 2.47(6×) Å. Second shell is very close to Ni-Ni shell of Ni fcc structure at 2.50(12×) Å and therefore the two phases are indistinguishable.

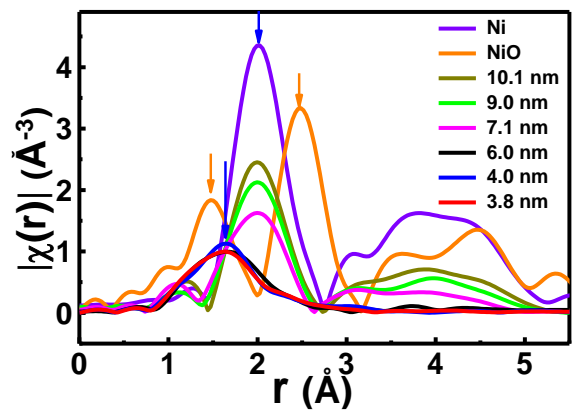

Figure 7 χ (r) vs r spectra of Ni Foil, NiO, 3.8 nm, 4.0 nm, 6.0 nm, 7.1 nm, 9.0 nm and 10.1 nm nanoparticle samples.

Consequently, experimental χ(r) spectra of 3.8 nm, 4.0 nm and 6.0 nm samples in which both Ni fcc and hcp phases exists, have been fitted from 1 to 2.5 Å assuming a Ni-O shell at 2.08(6×) Å and two Ni-Ni shells at 2.31(6×) Å and 2.47(6×) Å of the Ni hcp phase. Ni-O shell is due to the bonding of the surface atoms with the surfactant ligands as suggested by XANES analysis also. As the particle size decreases, proportion of surface atom increases and their contribution become significant. Hence, NiO-like non-metallic contribution is prominent in samples with smaller particle size. Since, we observe negligible oxidation but sign of coordination with surfactants [16], NiO-like contributions are attributed to these coordinated bonding.

For the samples with particle size larger than 6.0 nm, peaks observed in the experimental spectra confirm long range order in these samples. However, for samples with particle size less than 6.0 nm, only first shell has been observed. These samples also do not show any prominent Ni-Ni peak in spite of the XRD data confirms Ni metal lattice in all the samples. Absence of higher coordination shells may be attributed to the presence of multiple phases in these samples, which may lead to cancellation of the contribution of different paths. Figure 8 shows the experimental EXAFS spectra of Ni NP samples along with the best fitted theoretical plots carried out as above and the fitted results have been tabulated in Table S2. It can be seen from figure 8 that amplitudes of different peaks in EXAFS spectra of the Ni NPs are considerably less implying significantly low coordination numbers compared to bulk Ni as presented in table S2.

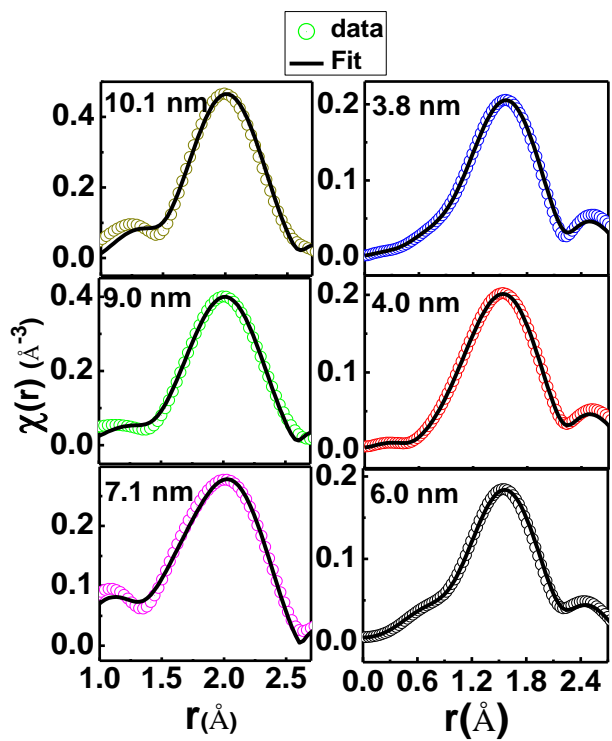

**Figure 8** Experimental χ(r) data (symbols) with best theoretical fits (continuous curves) of 3.8 nm, 4.0 nm, 6.0 nm, 7.1 nm, 9.0 nm and 10.1 nm nanoparticle samples.

Taking cue from XRD data analysis, they are ascribed to the presence of hcp phase along with fcc phase in these smaller NP samples wherein the contributions of surface atoms become significant in addition to ligands [16]. Since hcp and fcc phases only were not sufficient to best fit EXAFS data, NiO was also included in the fitting. This enabled the best fitting. Figure 7 shows significant contributions beyond the first coordination shell, observable in the r-space spectrum. This signature of long-range order is characteristic of a crystalline structure [35]. The considerably reduced amplitudes of spectral intensity of the Ni nanoparticles, compared to that of the bulk are usually attributed to nanosize effects [36-38]. Present 3.8 nm, 4.0 nm, 6.0 nm, 7.1 nm, 9.0 nm and 10.1 nm samples show reduction in EXAFS intensity of 61.7%, 62.6%, 61%, 34%, 24% and 20.5% in the *k*-space (Figure 6) and 77%, 74%, 77% 62.6%, 51% and 43.6% in the *r*-space (Figure 7) compared to those of bulk nickel. They are plotted in figure 9. They exhibit nearly a linear fall as the particle size increases. These reductions in intensities are consistent with earlier reports on different nanoparticle systems though with limited data therein [34,35,38-41]. For example, while it is about 8% theoretically as Moreau et al. [38] calculated for 5.5 nm Ni nanoparticles, they actually found 41% reduction in both k- and r-space for this sample experimentally. Comparable experimental values of 25% in k-space and 20% in r-space for 6 nm NiO NPs were reported by Anspoks et al. [36,37] While Cheng et al. [34] reported 70% reduction in the first shell peak amplitude of FT-EXAFS spectra of Co nanoparticles, Haubold et. al. [39,40] observed a 30-40% reduction in FT–EXAFS spectra for Cu nanoparticles and a ~50% reduction in the first peak of FT-EXAFS spectra in W nanoparticles. We ascribe such significant reduction in the EXAFS spectral intensity to the combined effects of (i) hcp and fcc lattice formation, (ii) probable P substitution [16,38], (iii) disorders in nanostructured system due to grain boundary and or (iv) coordination of ligands of surfactants to the surface atoms [16]. They result in the corresponding reduction in the coordination numbers compared to bulk counterpart as are evident from table S2.

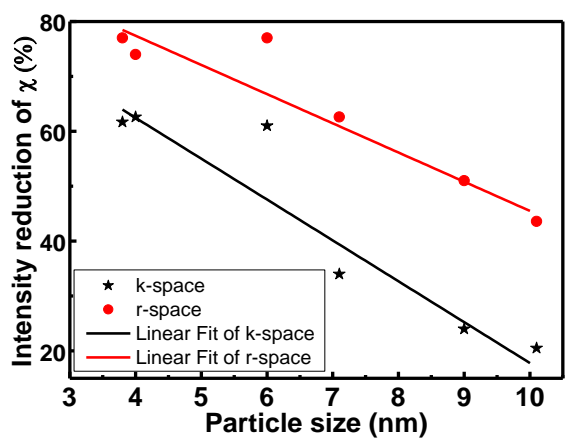

**Figure 9** The percentage reduction in the extended X-ray absorption fine structure intensity compared to that of Ni foil with increase in particle size. The k- and r- space data points (symbols) along with the best linear fits (curves) are shown in the legend.

Further, we have also tried to estimate the expected reduction in the first shell Ni coordination incorporating the effect of nanosize of particles following the method described by Calvin et. al. [41] assuming the nanoparticles to be spherical in shape and the expected coordination numbers have been given in Table S3. These coordination numbers are quite close to that of the theoretical value of 12.

This is in contrast to the much lower coordination numbers for Ni-Ni paths compared to that expected theoretically (Tables S2). This anomaly can only be explained by assuming the presence of some other phases that has not been considered appropriately during fitting. For example, since we do not have the idea of exact nickel compound formed, we have simply assumed that it was just NiO only. This however is not apparently true [16]. For samples with particle size less than 6.0 nm Ni fcc structure is also present along with Ni hcp structure. Due to the proximity of Ni fcc and Ni hcp structures, their percentage contributions couldn't be ascertained in the fitting with particle size less than 6.0 nm. This has been attributed to the dominant contribution from the significantly large number of surface atoms in the presence of TOP and oleylamine ligands on nanoparticles. These atoms therefore have enhanced (i) distortion in the local lattice structure due to coexistence of fcc and hcp lattices and (ii) covalent bonding of metal Ni with ligands and native oxygen atoms as size decreases. The present findings are in contrast to that of Illy et al. [42], who reported the direct evidence of size-dependent structural transition in nanosized nickel particles. However, unfortunately, since they characterized the samples with TEM using selected area electron diffraction only, they could identify hcp structure only for 3-4 nm particles and above this, only fcc structure, not the coexistence of fcc and hcp below 6 nm and existence of only fcc phase above it. These complicacies were evident from magnetization study also.

3.4 *Magnetization study*

Magnetization can be a powerful technique to study possible identification of structural phases in a multiphasic material for which the magnetization characteristics are different. This is applied in mixture of hcp and fcc structures of 8.5 nm and 18 nm Ni nanoparticles, leading to exhibition of two distinct zero-field cooled (ZFC) magnetization peaks due to the superparamagnetic blocking ($T_B$) at 12 K and 220 K, respectively [18]. They are further corroborated with ZFC magnetization peak at ~ 12 K of pure hcp Ni of 12.9 nm and above 300 K of pure fcc Ni of ~30nm [43]. Considering that these situations are applicable in the present samples, we assign double peaks in ZFC magnetization data (Figure 10) at 5 K and 200.9 K of 3.8 nm sample, 3.2 K and 117.4 K of 4 nm sample, less than 2 K and 81.9 K of 6.0 nm sample due to the presence of hcp and fcc phases in these smaller particle size samples. Note the systematic reduction of $T_B$ of hcp phase as the particle size increases from 3.8 nm to 6.0 nm through 4.0 nm. Relatively lower $T_B$ in the present samples, compared to those of distinct mixture of two phases or only hcp phase [18,43], is tentatively attributed to close proximity interaction with fcc phase and or related to the much smaller particle size for hcp phase in the naturally coexisting core-shell structure. In the larger particle size samples, only a single peak each in ZFC data is apparent supporting the finding from XRD analysis of pure fcc phase. Moreover, FC curves exhibit systematic evolution in the magnetic behaviours with increase in particle size. Thus, it is concluded that hcp and fcc phases coexist for particles with average size up to 6.0 nm beyond which only pure fcc phase triumphs. In order to understand better the magnetic properties of these nanoparticles, magnetization (M) as a function of magnetic field (H) was carried out at 3 K and 300 K. M-H loop is invisible for 3.8 nm sample at 300 K, while all others show M-H loops. They are indicative of ferromagnetic state at 300 K. Ferromagnetic state is attributed to the surface spin states that are normally found in nanoparticles [44] since $T_B$ for fcc phase exists well below 300 K (Figure 10). M-H loops are seen (Figure 11) however at 3 K for all the samples indicating that they are ferromagnetic at this temperature justifying superparamagnetic states above $T_B$ consistent with earlier reports on other nanoparticles [45]. Coercive fields $H_C$ determined from these M-H loops are plotted in Figure 11 (f).

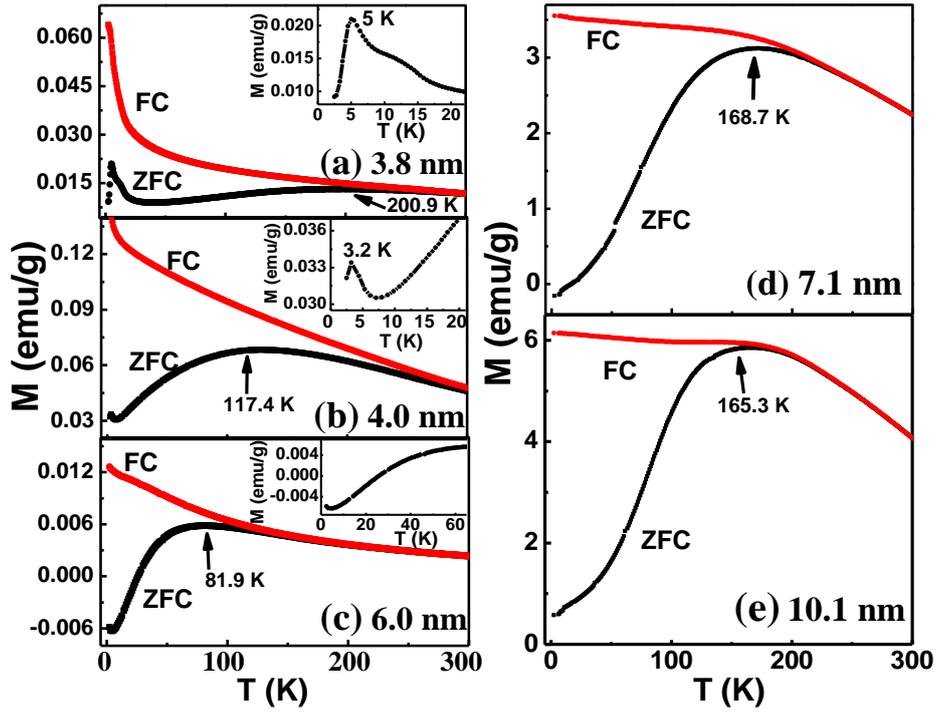

**Figure 10** Magnetization data at 50 G. Magnetization versus temperature of the Ni (a) 3.8 nm, (b) 4.0 nm, (c) 6.0 nm, (d) 7.1 nm and (e) 10.1 nm nanoparticle samples. Insets: low temperature regions of zero-field cooled data.

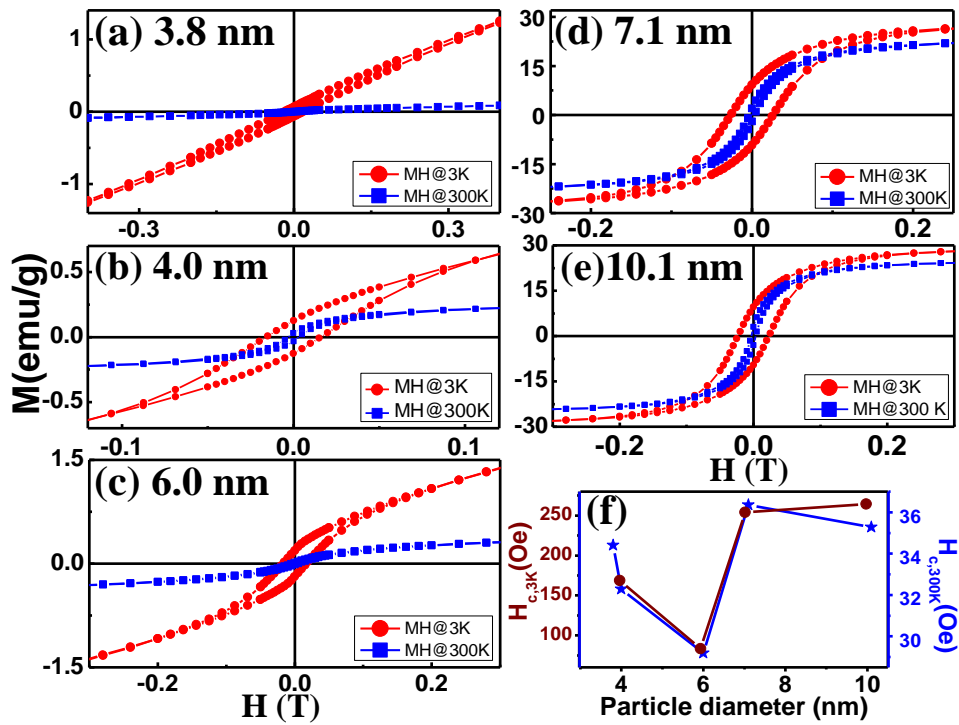

**Figure 11** Magnetization versus applied field (M-H) loops at 3 K and 300 K. Magnetization of the Ni (a) 3.8 nm, (b) 4.0 nm, (c) 6.0 nm, (d) 7.1 nm and (e) 10.1 nm nanoparticle samples. (f) Coercive field versus particle size at 3 K and 300 K

## 4. Conclusions

Detailed analysis of X-ray diffraction (XRD), synchrotron radiation X-ray absorption fine structure (EXAFS) and magnetization data of monodispersed nickel nanoparticles (NPs) of different sizes (3.8 nm to 10.1 nm) that exhibit natural nanolattices has been carried out. It has been demonstrated that particles of size in the range 3.8 nm to 6.0 nm have rather complex coexistent atomic hexagonal closed-packed and face-centred cubic crystal structures, and associated highly reduced coordination numbers, disturbed electronic structure, and surface atom coordination to the surfactants. They lead to complex magnetic properties of coexistent ferromagnetism and superparamagnetism. Trend is significantly different when the particle size is above 6.0 nm. They have been ascribed to dominant roles of the ligands of trioctylphosphine and oleylamine over the surface nickel atoms in forming covalent bonds and concomitant disturbance in the atomic lattices.

**Acknowledgments**
The Authors gratefully thank Dr. N. P. Lalla, Dr. M. Gupta and Dr. R. J. Choudhary, UGC-DAE CSR, Indore for TEM, XRD and SQUID-VSM data, respectively. This work is based on the routine research work carried out at UGC-DAE CSR, Indore, India. We are thankful to Director, Dr. A. K. Sinha and Centre-Director, Dr. V. Ganesan, CSR, Indore, India for their encouragements.

## Supplementary Data

# Size-induced structural phase transition at ~6.0 nm from mixed fcc-hcp to purely fcc structure in monodispersed nickel nanoparticles

Tarachand[1], Vikash Sharma[1], Jaiveer Singh[2], Chandrani Nayak[3], Dibyendu Bhattacharyya[3], Netram Kaurav[4], Shambhu N. Jha[3] and Gunadhor S. Okram[1]

[1] UGC-DAE Consortium for Scientific Research, University Campus, Khandwa Road, Indore 452001, MP, India
[2] Department of Physics, ISLE, IPS Academy, Rajendra Nagar, Indore 452012, MP, India
[3] Atomic & Molecular Physics Division, Bhabha Atomic Research Centre, Mumbai 400085, Maharastra, India
[4] Department of Physics, Government Holkar Science College, A. B. Road, Indore 452001, MP, India

**Synthesis of nickel nanoparticles** Thermal decomposition method was used to synthesize the nanoparticles as detailed in ref. [16]. Briefly, 1 ml (i.e. 2.24 mM) of preheated (215 °C) trioctylphosphine (TOP) (90% Aldrich) was added in the already degassed (at 100 °C for 30 min) solution of 1.02 g nickel acetylacetate (95% Aldrich) and 8 ml oleylamine (70% Aldrich). The resulting solution was further heated at 220 °C for 2 h under argon atmosphere. This led to black precipitate of nickel nanoparticles (NPs). Solution was then cooled to 27 °C, and centrifuged by adding ethanol (99.9% Jiangsu Huaxi) to extract and wash the nanoparticles. Washing was done four times. Similar procedures were followed for 2 ml, 3 ml, 5 ml, 8 ml, 10 ml and 12 ml of TOP at fixed (8 ml) OAm; their TEM size 10.1 nm, 9.0 nm, 7.1 nm, 6.0 nm, 4.0 nm and 3.8 nm, respectively, denoted these samples.

**X-ray Diffraction (XRD)** The Bruker D8 Advance X-ray diffractometer with Cu Kα radiation (0.154 nm) in the angle range 30-90° was used for laboratory method of XRD measurements of the samples in powder form; X-rays were detected using a fast counting detector based on silicon strip technology (Bruker LynxEye detector).

**Transmission electron microscopy (TEM)** Nanoparticle images were recorded using transmission electron microscopy (TECHNAI-20-G$^2$) by drop-casting the well-sonicated solution of a few milligrams of nanoparticles dispersed in about 5 ml ethanol on carbon-coated TEM grids. TEM was operated at 120 keV. At least 200 particles were analyzed per sample to obtain a representative size distribution.

**Sample preparation for X-ray absorption (XAS) experiment** The monodispersed Ni NP samples of appropriate weight, estimated to obtain a reasonable edge jump were taken in powder form. Each of them was mixed thoroughly with cellulose powder to obtain total weight of approximately 100 mg to make 2.5 mm thick homogenous pellets of 12.5 mm diameter using an automated hydraulic press. This is to avoid thickness effects (i.e., self-absorption [46] and satisfaction of the condition $\Delta\mu x \leq 1$ for the good absorption edge steps, where $\Delta\mu$ is the step in the absorption coefficient $\mu(E)$ value at the absorption edge and x is thickness of the sample pellet.

**XAS experiment** X-ray absorption measurements for Ni K edge of the monodispersed NC Ni samples in the above pellet forms were carried out at the scanning extended X-ray absorption fine structure (EXAFS) beamline (BL-9) at the INDUS-2 synchrotron radiation source (2.5 GeV, 100 mA) of the Raja Ramanna Centre for Advanced Technology, Indore, India [47]. The beamline covers a photon energy range of 4-25 KeV and has a resolution of 10,000 at 10 KeV. The data collection uses a double crystal monochromator (DCM) with a Rh/Pt coated cylindrical pre-mirror for vertical collimation of the beam and a similar post-mirror placed upside down for horizontal focusing of the beam. Second crystal of the DCM, which is a sagittal cylinder with radius of curvature in range of 1.28 - 12.91 meter, provides horizontal focusing to the beam. For measurements in transmission mode, sample was placed between two ionization chamber detectors. First detector measures the incident flux ($I_0$) and second detector measures the transmitted intensity ($I_t$). From these intensities the absorbance of the sample was determined as a function of energy. The X-ray absorption spectra of the samples for Ni K edge were recorded in the energy range 8280 - 8900 eV to cover both the X-ray absorption near-edge structure (XANES) and EXAFS regions. Five scans were obtained per sample using 15 minute scans at a step size of 1.0 eV.

**Fullprof Rietveld fitting of XRD**

The nearest-neighbor NN bond distance of 2.665 Å in hcp and 2.5095 Å in fcc nickel of literature value in bulk form comes to near 2.60 - 2.64 Å and 2.49 - 2.51 Å on nanostructuring in the present samples, showing significant change in the (local) crystal structure. Coordination number for both fcc and hcp is 12.

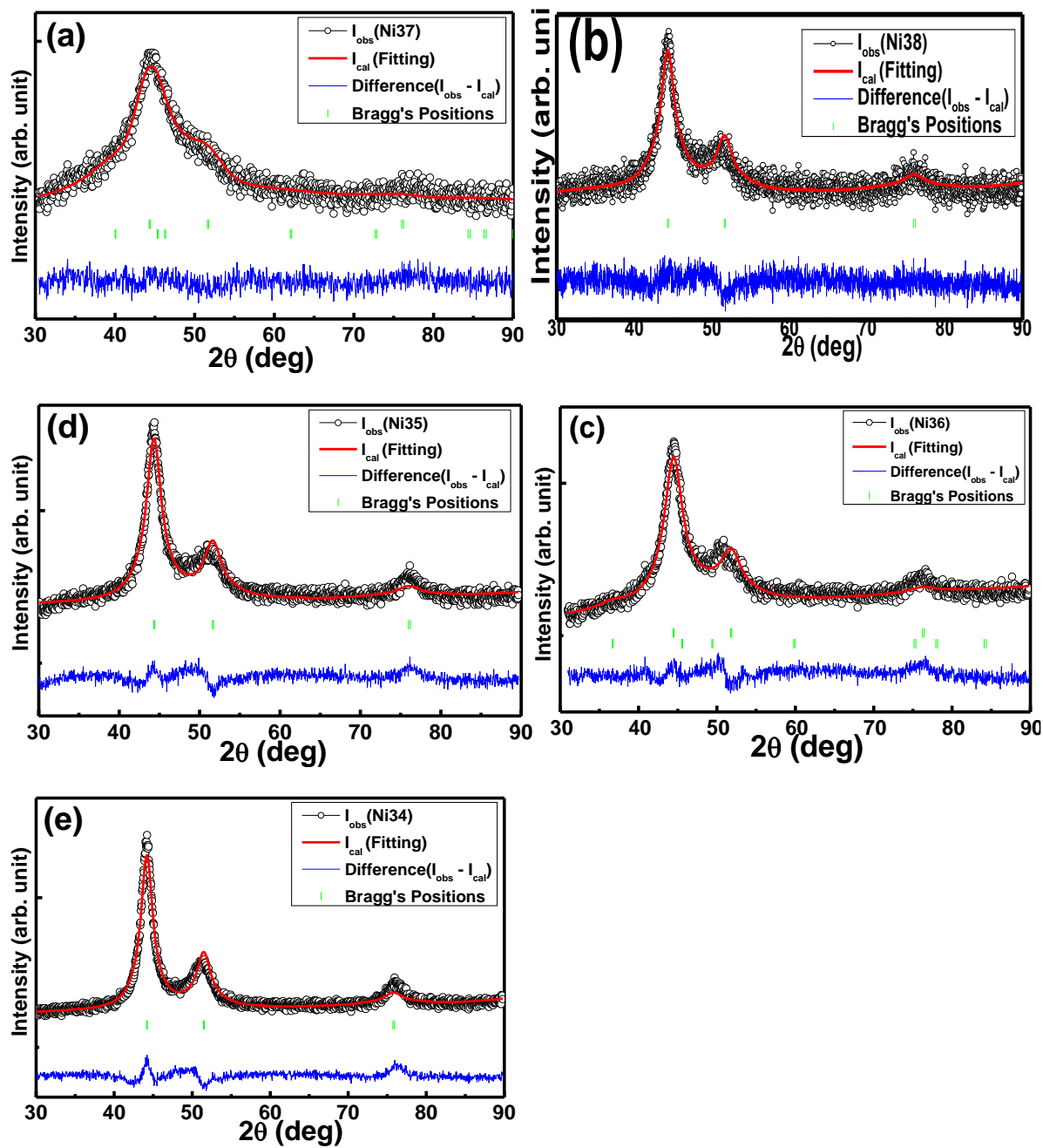

**Figure S1** XRD patterns of (a) 4 nm (Ni37), (b) 6.0 nm (Ni38), (c) 7.1 nm (Ni36), (d) 9.0 nm (Ni35) and (e) 10.1 nm (Ni34) nickel nanoparticle samples along with Rietveld fitted (red) curve, difference (blue) curve and Bragg peak positions (vertical lines) of fcc and hcp structure of nicke.

**Theory of XAS** When an x-ray is absorbed by a core-level electron with a binding energy, a photoelectron with wave number $k$ is created and knocks out of the absorbing atom. Absorption takes place only when the photoelectron state, with a quantum state at exactly the right energy, and also with the right angular momentum, is available. When there is no available state, there will be no absorption from that core level. Since the absorbing atom has its neighbors in the lattice, the photoelectron is scattered from the electrons of this neighboring atom, and the scattered photoelectron can return to the absorbing atom. While the absorption coefficient depends on whether there is an available electronic state at the right energy and momentum, presence of the photoelectron scattered back from the neighboring atom will alter the absorption coefficient.

X-ray absorption is thus a transition between two states: an initial state $\langle i|$ (having an X-ray photon, a core electron and no photoelectron) to a final state $|f\rangle$ (of no X-ray photon, a core hole and a photoelectron), and its coefficient $\mu(E)$ is described by the *Fermi's Golden rule* $\mu(E) \propto |\langle i|H|f\rangle|^2$, where H is the interaction term that comprises the quantized vector potential **A** of the theory of radiation. This can be reduced to a term that is proportional to $e^{ikr}$, where *k* is the magnitude of wave vector and r is the magnitude of position vector in real space of the photoelectron. The initial state is a tightly bound core-level that can be approximated by delta function, say a 1s level for atomic number Z extends to around $a_0/Z$, where $a_0$ is the Bohr radius (~0.529 Å). The change in final state is just the wave-function of the scattered photoelectron, $\Psi_{scatt}(r)$.

**Table S1** Lattice parameters obtained from Fullprof Rietveld fittings of XRD of samples

| Sample Name (Phase) | Wyckoff Position (x, y, z) | Thermal factor (β) | Occupancy | Phase Fraction (%) | Lattice parameter (Å) | Volume of unit cell (Å³) | Bond length (Å) |
|---|---|---|---|---|---|---|---|
| Ni31 [3.8 nm] (FCC) | 0, 0, 0 | 0 | 1 | 3.48 | a = b = c = 3.5330 | 44.099 | 2.498 |
| **Ni31 (HCP)** | 1/3, 2/3, 1/4 | 0 | 1 | 96.52 | a = b = 2.64, c = 4.22 | **25.471** | **2.640** |
| Ni37 [4.0 nm] (FCC) | 0, 0, 0 | 0 | 1 | 11.98 | a = b = c = 3.5390 | 44.324 | 2.502 |
| **Ni37 (HCP)** | 1/3, 2/3, 1/4 | 0 | 1 | 88.02 | a = b = 2.6, c = 4.0 | **23.417** | **2.600** |
| Ni38 [6.0 nm] (FCC) | 0, 0, 0 | 0 | 1 | 100 | a = b = c = 3.5413 | 44.411 | 2.5041 |
| Ni36 [7.1 nm] (FCC) | 0, 0, 0 | 1.222 | 1 | 100 | a = b = c = 3.5400 | 44.362 | 2.503 |
| Ni35 [9.0 nm] (FCC) | 0, 0, 0 | 09 | 1 | 100 | a = b = c = 3.5400 | 44.362 | 2.503 |
| Ni34 [10.1 nm] (FCC) | 0, 0, 0 | 1.5 | 1 | 100 | a = b = c = 3.5500 | 44.739 | 2.510 |

Scattering-path amplitude $f_i^{eff}(k)$ and phase $\delta_i(k)$ of the photoelectron are dependent on Z of the neighboring atom, where $i$ stands for $i^{th}$ shell. As a result, EXAFS is proportional to the amplitude of the scattered photoelectron at the absorbing atom. This consideration for one pair of absorbing atom and scattering atom is summed up in the total EXAFS from a sample to average over millions of atom pairs. Now, even if neighboring atoms are of the same type as absorbing atoms, the thermal and static disorder in the bond distances give a range of distances that will affect the EXAFS. This gives rise to the factor, $e^{-2k^2\sigma_i^2}$, where $\sigma_i^2$ is the mean-square deviation in the position of the $i^{th}$ atom at an effective half-path-length or the interatomic distance for single-scattering paths, $r_i$. In addition, other conduction electrons, phonons, and so on also scatter the outgoing photoelectron inelastically. This is described by $e^{-2r/\lambda_i(k)}$ with $\lambda_i(k)$ as mean free path of the photoelectron.

Thus, EXAFS, χ(k), with some algebra in the Fermi's Golden rule after taking into accounts of the factors described above turns out to be the sum of all contributions, $\chi_i(k)$, from groups of atoms that lie at approximately same distances from the absorbing atom at the $i$th shell, and is given by [48]

$$\chi(k) = \sum_i \frac{S_0^2 N_i}{kr_i^2} \left|f_i^{eff}(k)\right| \sin\left[2kr_i - \frac{4}{3}\sigma_i^{(3)}k^3 + \delta_i(k)\right] e^{-2\sigma_i^2 k^2} e^{-2r/\lambda_i(k)} \quad \ldots (1)$$

where $S_0^2$ is the passive electron reduction factor, $N_i$ is the coordination number, and $\sigma_i^{(3)}$ is the third cumulant. Equation 1 indicates that EXAFS is an inherently local probe due to $\lambda_i(k)$ term and $R^{-2}$ term, restricting within 5 Å or so from the absorbing atom. Second, EXAFS oscillations will consist of different frequencies that correspond to different lengths for each coordination shell.

**EXAFS analysis** The data processing was done using the IFEFFIT software package [49]. The AUTOBK code in IFEFFIT is used first to normalize the absorption coefficient, $\mu(E)$, and separation of the EXAFS signal, $\chi(E)$, from the isolated atom absorption background, $\mu_0(E)$ for each measured X-ray absorption spectrum. This gives the desired oscillations in the absorption spectra, whereby the energy dependent absorption coefficient $\mu(E)$ is converted to EXAFS function $\chi(E)$ or $\chi(k)$ defined as [5]

$$\chi(E) = \frac{\mu(E) - \mu_0(E)}{\Delta\mu_0(E_0)}$$

$$\text{or } \chi(k) = \frac{\mu(k) - \mu_0(k)}{\Delta\mu_0(k_0)}, \qquad (1)$$

where $E_0$ is energy at the absorption edge (chosen at the middle of the absorption edge jump), $k = \sqrt{2m(E-E_0)}/\hbar^2$ and m is the electron mass. Minimization of the signal in the low-$r$ region of the Fourier transformed $\chi(k)$ data was used to choose the background absorption function, $\mu_0(k)$. Then, the theoretical photoelectron scattering amplitudes and phase shifts were computed by fitting the experimental data with the theoretical spectra through the FEFF 6.0 code using a model structure. Finally, the theoretical EXAFS data was fitted to the experimental EXAFS data in $r$-space by Fourier transforming both the theoretical and experimental data.

We have carried out the EXAFS analysis following a model of fcc Ni for samples with larger particle size and mixture of fcc and hcp Ni for samples with smaller particle size along with NiO Ni hcp structure. Thus, the first path is assumed to consist of a path from fcc and a path from hcp structures each having coordination of 6 leading to a total coordination of 12. The final bond lengths are given in Table S2. The main peak in the FT-EXAFS spectra for the samples with larger crystallites appears at 2 Å. This matches Ni-Ni distance of 2.4-2.5 Å of bulk Ni. In case of samples with small crystallites, main peak appears at much lower value of 1.7 Å with broadening in larger values. To take care of this, we have taken Ni-Ni bulk distance along with a Ni-O path to represent any other compounds of Ni due to the ligands also [16].

**Table S2** EXAFS Fitted parameters

|  | Theoretical |  | 3.8 nm |  |  | 4.0 nm |  |  | 6.0 nm) |  |  |
|---|---|---|---|---|---|---|---|---|---|---|---|
| R Factor |  |  | 0.003 |  |  | 0.002 |  |  | 0.003 |  |  |
|  | r (Å) | N | r (Å) | N | $\sigma^2$ | r (Å) | N | $\sigma^2$ | r (Å) | N | $\sigma^2$ |
| Ni-O | 2.08 | 6 | 2.01(1) | 3.1(6) | 0.003(1) | 1.99(1) | 3.2(2) | 0.004(1) | 2.00(1) | 3.1(2) | 0.003(1) |
| Ni-Ni (fcc Ni) | 2.31 | 6 | 2.41(1) | 2.1(3) | 0.006(1) | 2.40(2) | 2.0(1) | 0.005(1) | 2.40(1) | 2.1(2) | 0.005(1) |
| Ni-Ni (hcp Ni) | 2.47 | 6 | 2.56(1) | 1.9(3) | 0.006(1) | 2.55(1) | 1.8(3) | 0.005(1) | 2.55(1) | 2.0(3) | 0.005(1) |

The fast decaying weak oscillations with $k$ of $\chi(k)$, derived from observed EXAFS data was weighted by $k$ or $k^2$ to amplify them at high values of $k$. The $\chi(k)k^2$ functions were then Fourier-transformed in $r$-space to generate the $|\chi(r)|$ versus $r$ (or FT-EXAFS) plots of the real space distances with respect to the absorbing atom. Since $\chi(k)$ is dependent on the coordination numbers N, bond distances $r_i$ and mean square deviations $\sigma_i^2$, we determined these parameters from the above analysis and studied various aspects related to the local structure of these monodispersed Ni NPs. X-ray absorption near-edge structure (XANES) was investigated to understand the type of bonding, type of ligand/ charge and symmetry of these NPs.

**Table S2** EXAFS Fitted parameters continued….

|  | Theoretical | | 10.1 nm | | | 9.0 nm | | | 7.1 nm | | |
|---|---|---|---|---|---|---|---|---|---|---|---|
| R Factor | | | 0.004 | | | 0.001 | | | 0.002 | | |
|  | r (Å) | N | r (Å) | N | $\sigma^2$ | r (Å) | N | $\sigma^2$ | r (Å) | N | $\sigma^2$ |
| Ni-O | 2.08 | 6 | - | - | - | - | - | - | 2.02(1) | 2.4(3) | 0.003(2) |
| Ni-Ni (fcc Ni) | 2.5 | 12 | 2.43(1) | 9.5(5) | 0.006(1) | 2.44(1) | 8.3(3) | 0.006(1) | 2.44(3) | 6.7(3) | 0.007(1) |

Table S3 Theoretical Coordination Numbers derived by the method developed by Calvin et al [41].

| Samples | Ni-Ni Co-ordination Number (N) | | |
|---|---|---|---|
| 10.1 nm | 11.78 | | |
| 9.0 nm | 11.75 | | |
| 7.1 nm | 11.69 | | |
|  | Coordination no. of 1st shell | Coordination no. of 2nd shell | Total Coordination |
| 6.0 nm | 5.78 | 5.77 | 11.55 |
| 4.0 nm | 5.64 | 5.62 | 11.26 |
| 3.8 nm | 5.05 | 4.99 | 10.04 |

**Note: All references are given in the end of main manuscript.**